\documentclass[10pt,conference]{IEEEtran}
\IEEEoverridecommandlockouts
\usepackage{cite}
\usepackage{amsmath,amssymb,amsfonts}
\usepackage{algorithmic}
\usepackage{algorithm2e}
\usepackage{graphicx}
\usepackage{textcomp}
\usepackage{multirow}
\usepackage{xcolor}
\usepackage{url}
\usepackage{multicol}
\usepackage{newunicodechar}
\usepackage{fontawesome5}
\def\BibTeX{{\rm B\kern-.05em{\sc i\kern-.025em b}\kern-.08em
    T\kern-.1667em\lower.7ex\hbox{E}\kern-.125emX}}
\begin{document}

\title{Establishing Traceability between Release Notes \& Software Artifacts: Practitioners' Perspectives}

\author{\IEEEauthorblockN{Sristy Sumana Nath}
\IEEEauthorblockA{\textit{Department of Computer Science} \\
\textit{University of Saskatchewan}\\
sristy.sumana@usask.ca}
\and
\IEEEauthorblockN{Banani Roy}
\IEEEauthorblockA{\textit{Department of Computer Science} \\
\textit{University of Saskatchewan}\\
banani.roy@usask.ca}

\and
\IEEEauthorblockN{Munima Jahan}
\IEEEauthorblockA{\textit{Department of Computer Science} \\
\textit{Thompson River University}\\
mjahan@tru.ca}}

\maketitle

\begin{abstract}
Maintaining traceability links between software release notes and corresponding development artifacts, e.g., pull requests (PRs), commits, and issues, is essential for managing technical debt and ensuring maintainability. However, in open-source environments where contributors work remotely and asynchronously, establishing and maintaining these links is often error-prone, time-consuming, and frequently overlooked. Our empirical study of GitHub repositories revealed that 47\% of release artifacts lacked traceability links, and 12\% contained broken links.
To address this gap, we first analyzed release notes to identify their What, Why, and How information and assessed how these align with PRs, commits, and issues. We curated a benchmark dataset consisting of 3,500 filtered and validated traceability link instances. Then, we implemented LLM-based approaches to automatically establish traceability links of three pairs between release note contents \& PRs, release note contents \& PRs and release note contents \& issues. By combining the time proximity feature, the LLM-based approach, e.g., Gemini 1.5 Pro, achieved a high Precision@1 value of 0.73 for PR traceability recovery. To evaluate the usability and adoption potential of this approach, we conducted an online survey involving 33 open-source practitioners. 16\% of respondents rated as very important, and 68\% as somewhat important for traceability maintenance.
\end{abstract}

\begin{IEEEkeywords}
Release notes, Traceability, GitHub, Human study
\end{IEEEkeywords}

\section{Introduction}

Maintaining accurate traceability links between release notes and their corresponding development artifacts, e.g., pull requests (PRs), commits, and issues, is fundamental for ensuring software quality and sustainability \cite{RelContentHuman22}. Release notes are primarily intended for end users, helping them decide whether it is worthwhile or safe to install or upgrade a new release \cite{arena1}. For libraries and frameworks, developers often act as end users and benefit from traceable release documentation.
Release notes also serve as a bridge, connecting change summaries to underlying commits, PRs, and issues.
Making these links explicit increases transparency, reduces documentation debt, and supports project onboarding \cite{icse24posterTRL}.
Thus, traceability benefits end users most directly, while indirectly helping maintainers, reviewers, and contributors.

Despite its importance, traceability in open-source projects is often overlooked, as contributors work voluntarily and across distributed time zones without formal process support. As a result, release notes are frequently only partially linked or not linked at all to development artifacts. Our empirical investigation into several popular GitHub repositories highlights the magnitude of this problem: 47\% of release artifacts lacked any form of traceability links, while an additional 12\% contained broken or incorrect links. Missing links make it difficult for developers to trace changes back to their source, slowing debugging, maintenance, and knowledge transfer \cite{traceabilityartifactmanagesys}. This gap also increases documentation debt, reduces trust in project communication, and hampers collaboration in open-source development \cite{SLRTDandTDM}.

Our study addresses this gap by exploring automated techniques to recover traceability links between release notes and development artifacts in open-source GitHub projects.
We began by systematically analyzing release notes from various GitHub repositories, focusing on categorizing their content into \textit{What}, \textit{Why}, and \textit{How} components. We observed that these elements are typically distributed across PR titles, commit messages, and issue descriptions. For example, the What is often described in the PR title, the Why in the issue discussion, and the How in the commit message. This observation formed the basis of our hypothesis that automated methods, leveraging both textual similarity and contextual features, could effectively recover these missing links.

\begin{figure*}
\centering
\includegraphics[width=6in]{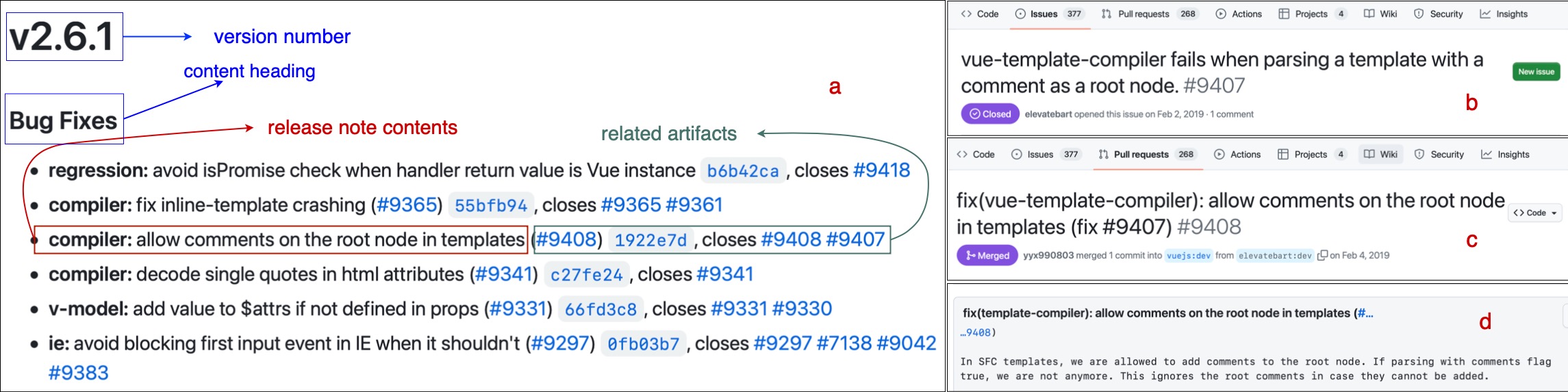}
\caption{Example of natural language artifacts: a) Release notes, b) Issues, c) Pull requests, and d) Commits}
\label{fig:Example_RN_PR_Iss_CM}
\end{figure*}

To test this hypothesis, we curated a benchmark dataset comprising 3,500 validated traceability link instances, drawn from a diverse set of GitHub projects and covering a wide range of programming languages, project sizes, and domains. We carefully filtered this dataset to remove ambiguous or broken links, ensuring a high-quality resource for evaluating traceability recovery methods. Our approach combined natural language processing (NLP) techniques and machine learning models, focusing particularly on large language models (LLMs) such as Meta LLaMA 3.1 and Gemini 1.5 Pro. These models were fine-tuned to predict traceability links based on both textual features (e.g., release note content, PR titles, commit messages, issue titles) and non-textual features (e.g., time proximity based on merge dates, commit dates, and issue resolution dates).

Our results demonstrate that LLM-based approaches significantly outperform traditional baselines, such as TF-IDF with cosine similarity or simple embedding-based methods. For instance, when time proximity signals were incorporated alongside textual features, Meta LLaMA 3.1 achieved Precision@1 values of 0.68 for PR recovery and 0.63 for issue recovery. Gemini 1.5 Pro further improved these results, achieving Precision@1 values of 0.73 for PRs and 0.70 for issues. These models also demonstrated strong Mean Reciprocal Rank (MRR) performance, indicating their robustness in ranking relevant artifacts near the top of the candidate list. The integration of time proximity proved particularly valuable, as it allowed the models to disambiguate between similar artifacts by prioritizing those temporally aligned with the release.

To complement our technical evaluation, we conducted a survey of 33 practitioners drawn from both open-source and enterprise software development communities. The survey aimed to understand current traceability practices, perceived importance of traceability, and openness to adopting automated solutions. Our findings reveal that while 16\% of respondents considered traceability maintenance very important, and 68\% somewhat important, the actual practice of linking artifacts to release notes was inconsistent. A significant proportion of participants (81\%) reported linking artifacts only sometimes, with 10\% rarely and 9\% never doing so. The main barriers identified included the manual effort required, the risk of introducing errors, and the lack of integrated tooling. Above 90\% of participants expressed strong interest in automated solutions that could help to trace the release note with their related artifacts. 

Our contributions through this study are fourfold. First, we present the first large-scale empirical study that quantifies the extent of traceability gaps between release notes and development artifacts in open-source repositories. Second, we release a high-quality benchmark dataset of 3,500 validated traceability links, designed to support future research in this area. Third, we propose and rigorously evaluate an LLM-based approach that recovers traceability links with high precision by leveraging both textual and non-textual features. Finally, we provide practitioner insights into the adoption of automated traceability solutions, highlighting the need for accuracy, explainability, and seamless integration with existing workflows.

\section{Motivating Example}\label{sec:motivatingex}

Consider a motivating example (shown in Fig. \ref{fig:Example_RN_PR_Iss_CM}) from the release notes of version 2.6.1 of an open-source project, where bug fixes are documented under the content heading ``Bug Fixes.” 
Fig. \ref{fig:Example_RN_PR_Iss_CM}b, c, and d showcase the issues, PRs, and commits of the selected content (red mark). These features enable developers to swiftly navigate through the changes with ease.
This enables contributors to trace the change back to the corresponding code and issue, helping them understand the context and the specific problems addressed.

For contributors, traceability clarifies the history and rationale behind changes, helping them avoid duplication, target fixes, and align with project goals. Manually maintaining these links is time-consuming, especially for those juggling multiple tasks or new to the repository. Automated traceability recovery reduces this burden, allowing contributors to focus on writing high-quality code.

For newcomers, traceability supports onboarding by showing how features and fixes evolved across release notes, commits, and issues. Clear links make it easier to navigate the codebase, identify opportunities to contribute, and avoid mistakes. AI-powered tools can make these connections more accessible, helping newcomers integrate smoothly and contribute effectively.

For reviewers, traceability provides critical context during code reviews, connecting PRs and commits to the issues they address and their representation in release notes. Without it, reviewers risk overlooking important details or misjudging changes. Automated traceability tools streamline this process, enabling faster, more informed reviews and protecting code quality.

\section{Methodology}\label{sec:method}
\begin{figure}
    \centering
    \includegraphics[width=3.7in]{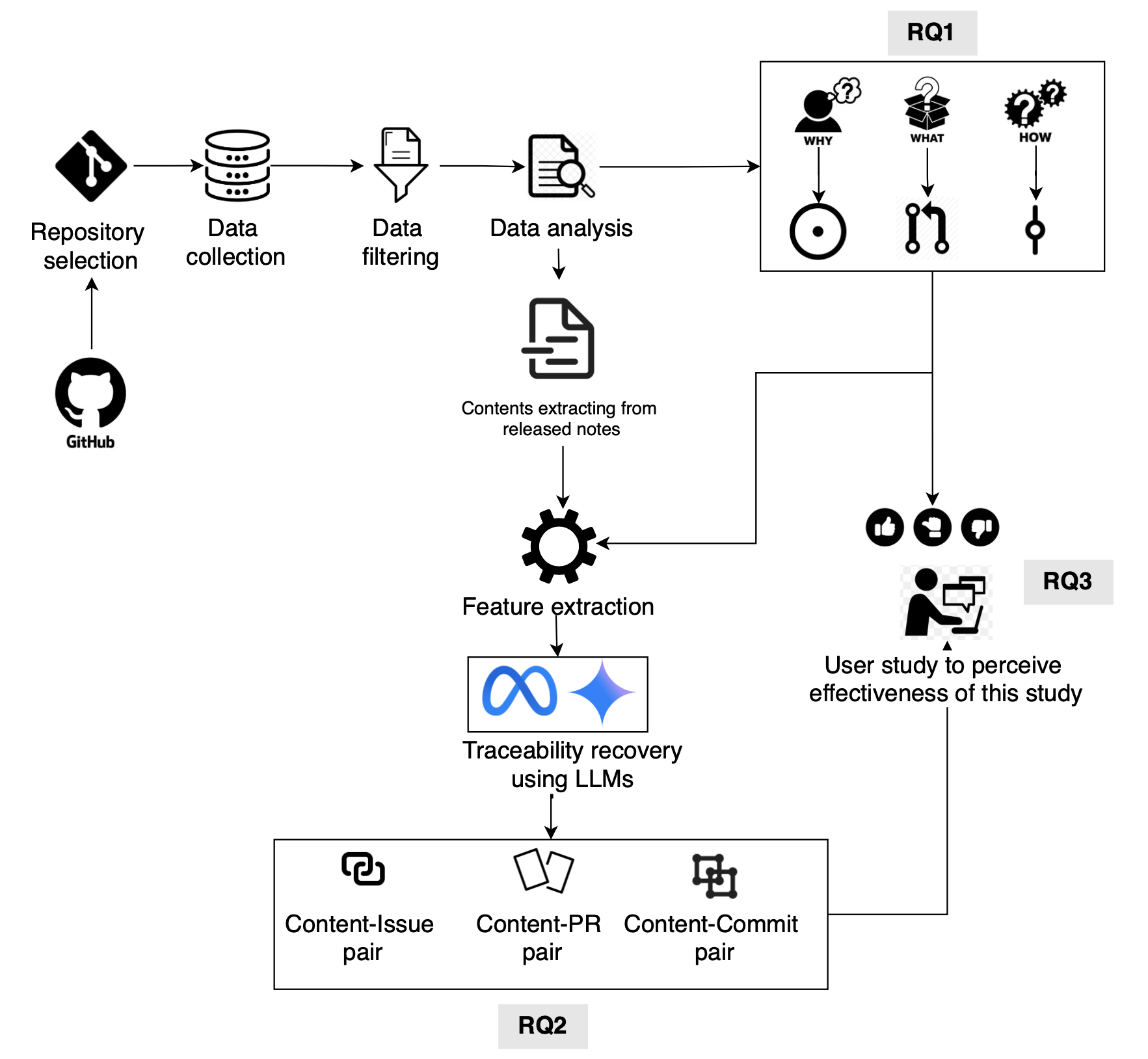}
    \caption{Research overview}
    \label{fig:overview}
\end{figure}

Our goal is to investigate whether the release notes contain \textit{What} information, \textit{Why} it needs to change and \textit{How} to use the information. Then we aim to propose an LLM-based model to generate release notes with include \textit{why} and \textit{how} information. By exploring the state of the art, we find previous techniques \cite{releaseseke2021, arena1, deeprelease} generate release notes based on the  \textit{What} information. Fig. \ref{fig:overview} shows an overview of our methodology,

\subsection{Research Questions}
Our study formulates the following research questions:
\paragraph{RQ1. What insights can we gain by analyzing release notes that link to their corresponding PRs, commits, and issues to enhance traceability?}
This research question aims to explore what valuable information can be uncovered through the analysis of release notes in open-source projects. By investigating how these notes reference or connect to pull requests, commits, and issues, we seek to understand their current role in supporting traceability. This will help identify patterns, gaps, and best practices for improving documentation and artifact linkage within software development processes.
\paragraph{RQ2. Can LLM-based approaches accurately recover traceability links?}
This research question focuses on evaluating whether an LLM-based method can accurately recover traceability links between release notes and their related pull requests, commits, and issues. This study intends to assess the effectiveness of LLMs in automating traceability link recovery and determine their potential to reduce manual effort, improve documentation quality, and support software maintenance tasks across open-source development environments.
\paragraph{RQ3. How do practitioners perceive the impact of the proposed traceability links between release notes and related software artifacts?}
This research question aims to gather and analyze practitioners’ opinions on the usefulness and impact of establishing traceability links between release notes and associated development artifacts. Through the survey, we seek to understand their perspectives on how such traceability affects collaboration, maintenance, debugging, and documentation quality. This will provide insights into real-world challenges, benefits, and desired improvements in traceability practices within software projects.

\begin{figure*}
    \centering
    \includegraphics[width=6in]{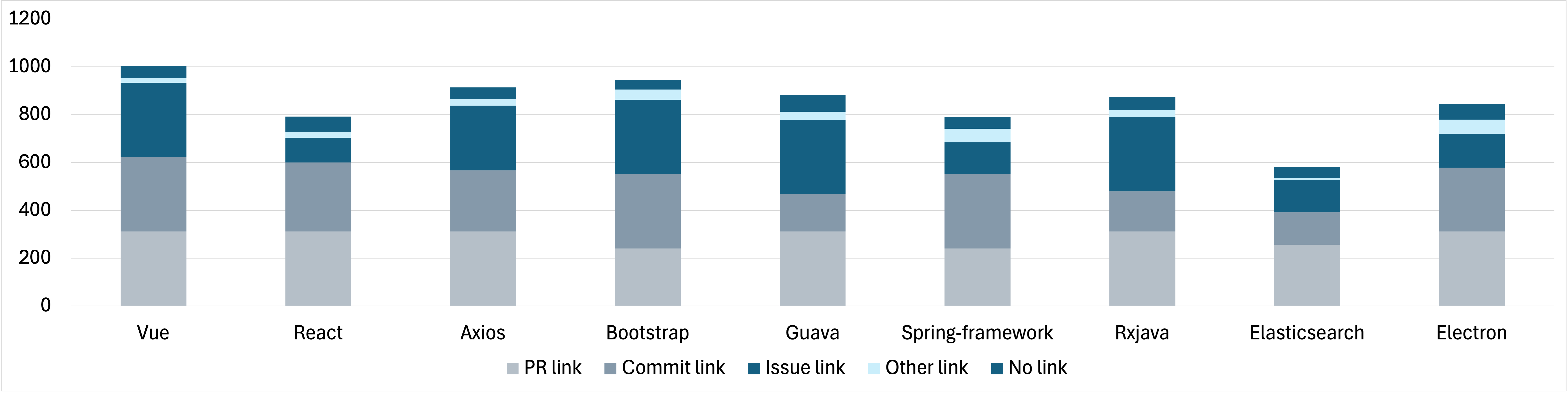}
    \caption{Dataset Overview}
    \label{fig:dataset}
\end{figure*}

\subsection{Project Selection and Data Extraction}

We began by collecting 10,520 release notes from open source GitHub repositories. To ensure maturity, popularity, and relevance, we focused on three widely used programming languages: JavaScript, Java, and Python. Projects were selected using GitHub’s search interface, sorted by stars, and filtered with the following criteria: (1) created at least three years ago and still active, (2) $\geq$1,000 forks and $\geq$8,000 stars, (3) $\geq$30 contributors, (4)$\geq$2,000 commits, (5) $\geq$500 resolved issues, (6) $\geq$500 merged pull requests, and (7) $\geq$30 published release notes. Applying these thresholds yielded 10 repositories (Fig. \ref{fig:dataset}).

From the selected repositories, we extracted release notes, commit messages, merged PR titles, and closed issue titles using the GitHub API. Along with the textual content, we collected metadata such as project name, version number, release date, commit date, commit URL, PR merge date, PR URL, issue closure date, and issue URL. Release note contents were segmented by release version to allow fine-grained alignment with related artifacts. This ensured that our dataset captured both the natural language used in release documentation and the associated technical artifacts.

\subsection{Data Preparation and Labeling}
For data preparation, we collected textual content from multiple artifacts, including release notes, commit messages, merged pull request titles, and closed issue titles. Each release note was segmented into individual sentences and tracked by its corresponding release version number in the dataset. As illustrated in Fig. \ref{fig:Example_RN_PR_Iss_CM}a, the release note sentences are highlighted in red, while Fig. \ref{fig:Example_RN_PR_Iss_CM}b, c and d show the associated artifact identifiers and hyperlinks that connect release notes to commits, PRs, and issues. In this way, we organized and structured the data for traceability analysis.

To construct a reliable ground truth, we focused only on release notes that contained explicit links to related artifacts. Release notes without links were excluded. When extracting release notes, we also captured the embedded hyperlinks, which allowed us to implement a rule-based labeling approach inspired by prior work on release note analysis \cite{RelContentHuman22}. Specifically:
\begin{itemize}
\item Hyperlinks containing $/commit/$ were labeled as links to commits.

\item Hyperlinks containing $/pull/$ were labeled as links to PRs.

\item Hyperlinks containing $/issues/$ were labeled as links to issues.
\end{itemize}

This procedure provided precise automatic labels between release note contents \& artifact pairs. From these candidates, we curated a dataset of 3,500 contents \& artifact links as ground truth. By leveraging the explicit linking practices already present in GitHub release notes, our dataset construction approach avoids subjective manual annotation and ensures reproducibility. The resulting dataset serves as a high-quality benchmark for evaluating automated traceability recovery methods.
Data and code are uploaded as additional materials.\footnote{https://github.com/sristysumana/TraceabilityStudyCascon2025}.

\subsection{Recover Traceability Using LLMs}

We implemented {Meta LLaMA 3.1} and {Gemini 1.5 Pro} to recover traceability links between release-note contents and their corresponding issues, pull requests (PRs), and commits. These models were chosen because they are state-of-the-art instruction-tuned large language models (LLMs) capable of reasoning over technical text (including code-related descriptions) and capturing semantics that traditional IR baselines (e.g., TF-IDF or simple Seq2Seq classifiers) often miss. Both models are accessible via public APIs, Hugging Face for LLaMA 3.1 and Google AI Studio for Gemini 1.5 Pro, supporting reproducibility and scalability in applied software engineering research.

For our study, we used the {Meta-LLaMA-3-8B-Instruct} model through the Hugging Face Inference API. Each request paired a single release-note sentence with a single candidate artifact text (issue title, PR title, or commit message) in a structured, binary prompt. We enforced deterministic decoding (\texttt{temperature = 0.0}, \texttt{max\_new\_tokens = 50}) so identical inputs yield identical outputs. The model returned concise judgments (\textit{``YES''} or \textit{``NO''}, sometimes with a brief explanation), which we mapped to a text score (\textit{YES}$\rightarrow1.0$, \textit{NO}$\rightarrow0.0$). This setup leverages LLaMA’s instruction tuning to reason effectively about code-adjacent natural language while keeping token windows small and the evaluation transparent.

{Gemini 1.5 Pro}  is a multimodal, instruction-tuned model designed for complex reasoning, code understanding, and factual tasks. We accessed it via Google AI Studio’s REST API, using the same prompt structure and deterministic, single-shot generation. Gemini’s extended context window and pretraining on technical corpora help align release-note content with commits, PRs, and issues. Importantly, time proximity was not included in the prompt; instead, we combined the LLM’s binary text score with a post-hoc temporal score measuring the distance between the release date and candidate artifact date, using a weighted fusion:

\begin{equation}
FinalScore = \alpha \cdot TextScore + (1-\alpha) \cdot TimeScore
\end{equation}

where $Window = 30$ days. We set $\alpha = 0.7$ to give higher weight to textual similarity while still incorporating temporal alignment, consistent with prior work emphasizing text as the dominant signal with context as a complementary feature \cite{linkcodeDoc, issuecommiticsme2021}. In our experiments, we compared results using only text similarity and using the fused text+time scoring.

\subsection{Baselines}
\textbf{TF-IDF + Cosine Similarity.}
This baseline \cite{icse24posterTRL} represents a traditional information retrieval approach. Both the release note contents and the candidate artifacts (pull request, commit, or issue title) are converted into TF-IDF (Term Frequency–Inverse Document Frequency) vectors. Cosine similarity is then computed between these vectors to quantify textual relatedness.

\textbf{Seq2Seq.} Sequence-to-Sequence neural model employs a sequence-to-sequence encoder-decoder architecture to learn relationships between release note contents (input sequence) and corresponding artifact descriptions (output sequence).

The proposed model captures deep semantics, context, and auxiliary information (e.g., time proximity); adaptable to multi-repository data.

\subsection{Evaluation Metrics}
In this study, we evaluate the traceability link recovery performance using Precision@1 and Mean Reciprocal Rank (MRR).

Precision@1 measures how often the top-ranked artifact retrieved by our model corresponds to the correct link. This is critical in practice, as developers tend to inspect only the first result when navigating release content links. It tells how often the model identifies the single most relevant artifact (issue/PR/commit) for a release note at first glance. This is crucial in traceability because users typically review the top suggestion.

\begin{equation}
\scriptsize
    \text{Precision@1} = \frac{\text{\# of correct top-1 predictions}}{\text{total \# of queries}}
\end{equation}

 MRR measures how highly the correct artifact is ranked, on average, across all queries. These metrics are widely adopted in traceability link recovery and information retrieval research, providing reliable indicators of model utility in real-world development scenarios.
\begin{equation}
\scriptsize
\text{MRR} = \frac{1}{|Q|} \sum_{i=1}^{|Q|} \frac{1}{\operatorname{rank}_i}
\end{equation}
Where $Q$ is the set of queries (e.g., release contents are linking) and $i$ is the position (rank) of the first correct artifact for query $i$.
MRR gives credit even if the correct link is not at the top, but is near the top. This reflects practical scenarios where developers skim the first few results. It is a standard in information retrieval and traceability studies.
\subsection{Participants Requirement for Online Survey}

\begin{figure}
    \centering
    \includegraphics[width=3in]{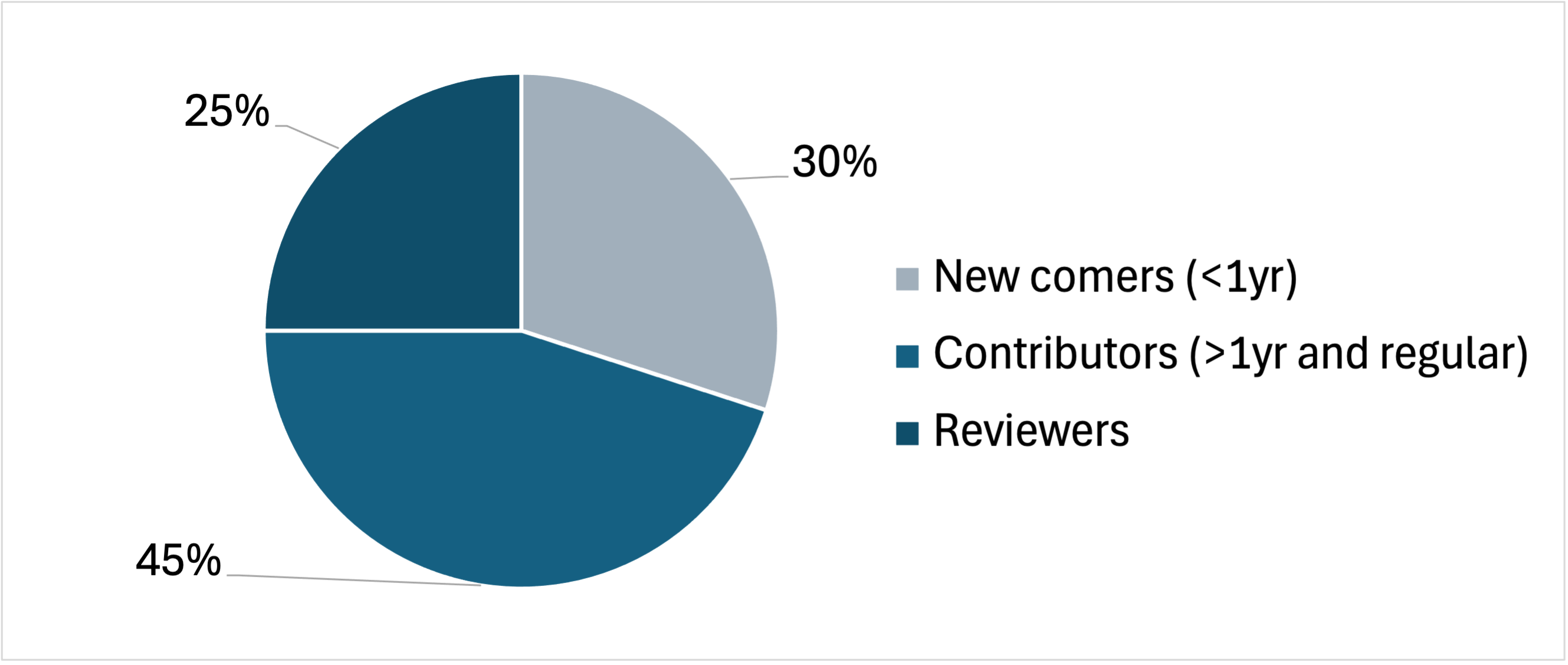}
    \caption{Participants' Role in GitHub}
    \label{fig:participants}
\end{figure}

The online survey for the traceability study is designed to gather insights from GitHub contributors regarding their practices for linking release notes to corresponding development artifacts. The survey included sections on participant demographics, GitHub activity, and their methods for tracking changes between release notes and development artifacts. It explored their understanding of traceability, the challenges they encounter, and their openness to adopting AI-based tools for improving traceability link recovery. The survey also assessed the effectiveness of current tools and practices, as well as the pain points in maintaining accurate traceability. Participants were recruited through targeted outreach on GitHub, utilizing repositories, issues, and discussions related to open-source development. Additional recruitment was conducted by sending emails where active open-source contributors. 
To recruit participants for our study on community smells in software development, we adopted a method commonly used in software engineering literature: directly contacting contributors to open-source projects via publicly available email addresses on GitHub. This approach has been effectively used in prior research to engage practitioners actively involved in real-world software projects. We identified contributors from several active and widely used GitHub repositories by analyzing their commit history and contributions. Personalized invitation emails were sent to these contributors, informing them about the study's objective and providing a link to the survey. Participation was entirely voluntary, and respondents were assured of anonymity and data confidentiality.
Fig. \ref{fig:participants} represents the percentage of newcomers (below 1 year of experience), contributors (more than one year of experience) and reviewers of GitHub.

\section{Results and Discussion}\label{sec:result}
\subsection{RQ1: What insights can we gain by analyzing release notes that link to their corresponding PRs, commits, and issues to enhance traceability?}\label{sec:rq1result}
Previous studies have indicated that release note contents are associated with three software artifacts in GitHub: commits (19\%), pull requests (32\%), and issues (29\%) \cite{RelContentHuman22}. Fig. \ref{fig:Example_RN_PR_Iss_CM} illustrates this relationship. Our study explored the structural patterns of release notes and found what, why, and how changes were included there.

In this study, we set out to investigate how release notes in open-source GitHub projects document the nature of software changes and how they can support traceability to development artifacts. Specifically, we examined the inclusion of What, Why, and How information in release notes, as these categories provide the key context users and developers need to understand, assess, and adopt software updates. Our analysis focused on assessing both the standalone presentation of this information in release notes and the extent to which linked PRs, commits, and issues help bridge informational gaps.

\begin{figure}
    \centering
    \includegraphics[width=2in]{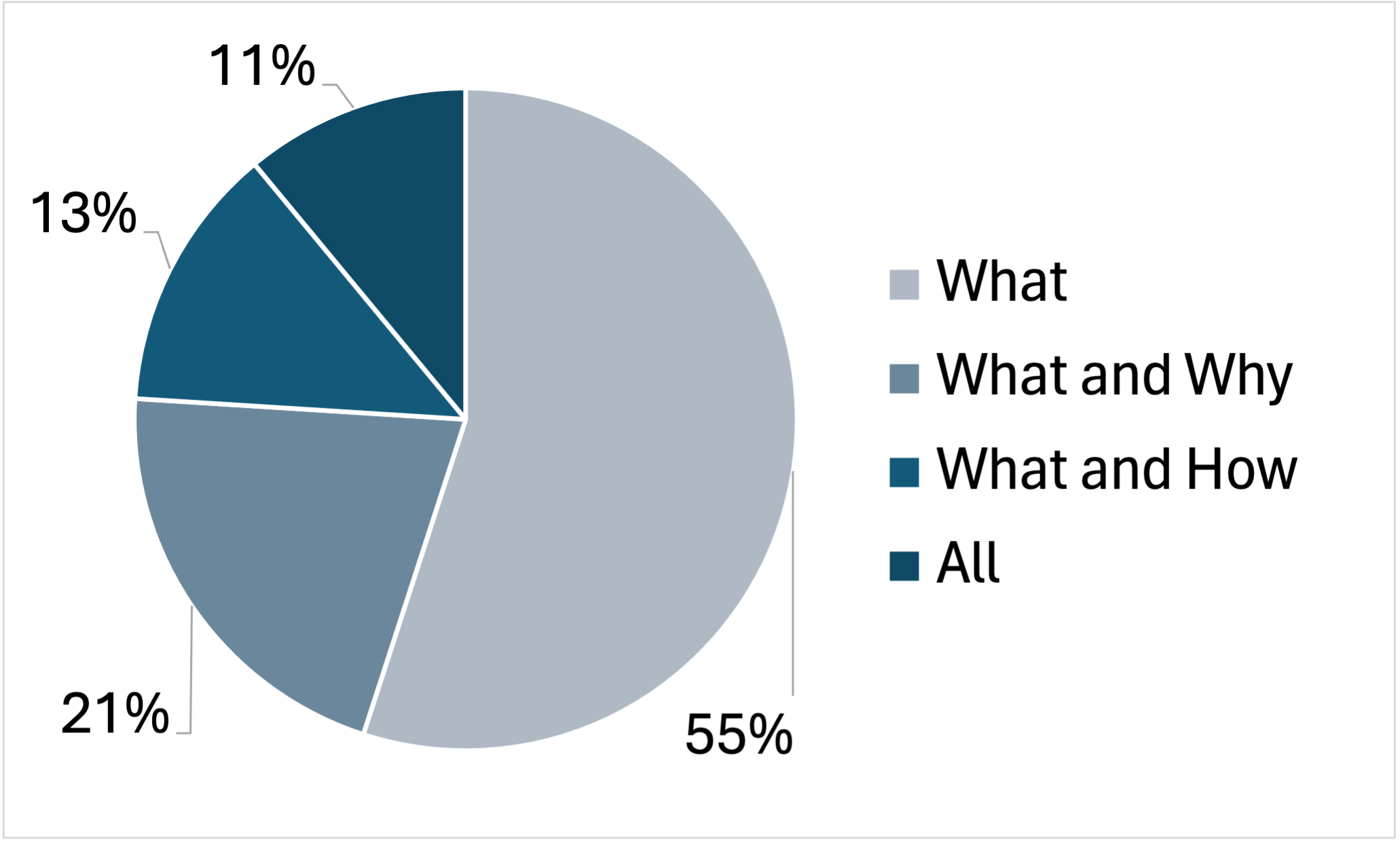}
    \caption{Percentage of containing What, Why and How information}
    \label{fig:piechartDiss}
\end{figure}

We categorized the presentation structures identified through manual analysis and discussed whether they included information on What, Why, and How changes were made. Our findings showed significant variation in how comprehensively release notes describe changes. Fig. \ref{fig:piechartDiss} summarizes the percentage distribution of release notes by the types of information they contain. 55\% of release notes contained only \textit{What} information. These release notes describe the features added, bugs fixed, or functionalities updated. However, they did not provide explanations for why the changes were made or how they were implemented. For example, a release note might state, ``Added support for nullable embedded entities," without further elaboration. 21\% of the release notes contained both \textit{What} \textit{and} Why information.
These release notes not only describe what was changed but also offer some motivation or rationale, such as addressing a known issue or fulfilling a user request.
13\% of release notes contained What and How information.
Here, the release notes described the changes and included some technical details about the implementation or approach used.
Only 11\% of release notes integrated all three types of information e.g., \textit{What}, \textit{Why}, and \textit{How}.
These were relatively rare and can be considered well-structured, providing a complete picture of the change’s purpose, description, and implementation.
The distribution highlights a heavy emphasis on documenting \textit{What} was done, with far less attention given to \textit{Why} and \textit{How}. This imbalance limits the usefulness of release notes as a standalone source of traceability and context.

To illustrate how release notes and linked artifacts can collectively provide comprehensive change documentation, we present an example from the typeorm/typeorm repository (Table \ref{tab:whatwhyhowinfo}). Our analysis offers several important insights into how release notes support traceability:

\faLightbulb\ Release notes predominantly document \textit{What} information.
The majority (55\%) of release notes describe only the changes made without justifying why the change was needed or how it was implemented. This aligns with practices in many open-source projects where release notes prioritize brevity and user-facing descriptions.

\faLightbulb\ Inclusion of \textit{Why} and \textit{How} is limited.
Only a small fraction (11\%) of release notes contained all three types of information. When the \textit{Why} information was present, it usually appeared in the form of references to issues (e.g., ``Fixes \#1234") or brief problem statements. How information, when provided, was usually in technical notes or references to commits or PRs.

\faLightbulb\ Linked artifacts can fill informational gaps.
We observed that linked pull requests effectively provide \textit{What} information through titles and descriptions, issues provide the \textit{Why}, and commits offer the \textit{How}. This pattern suggests that creating or recovering traceability links could enhance the overall documentation quality without overloading the release notes themselves.

\faLightbulb\ Trade-off between completeness and usability.
Release notes that try to document all What, Why, and How information directly tend to become long and harder to read. This could lead to reduced usability and lower adoption of the notes by users. Instead, linking to PRs, issues, and commits can allow users to explore deeper levels of detail as needed.

\faLightbulb\ Opportunity for automated traceability recovery.
Given that PRs, issues, and commits together contain What, Why, and How information, automated recovery of traceability links (using methods like LLMs) could offer a scalable way to provide well-structured release documentation. This approach would reduce documentation debt while keeping release notes concise.

\begin{table}[]
    \centering
    \scriptsize
        \caption{`What', `Why' and `How' information}
    \label{tab:whatwhyhowinfo}
    \begin{tabular}{p{29em}}\hline
         \textbf{Release note contents:} nullable embedded entities (\#10289) (e67d704) \\\hline
         \textbf{Issue title:} Cannot set embedded entity to null. (\textit{\textbf{Why}})\\\hline
         \textbf{PR title:} feat: nullable embedded entities. (\textit{\textbf{What}})\\\hline
         \textbf{Commit msg.:} feat: nullable embedded entities. fix: ignore embedded columns if not selected". (\textit{\textbf{How}})\\\hline
         		
    \end{tabular}
\end{table}

\subsection{RQ2: Can LLM-based approaches accurately recover traceability links? }

\begin{figure*}
    \centering
    \includegraphics[width=7in]{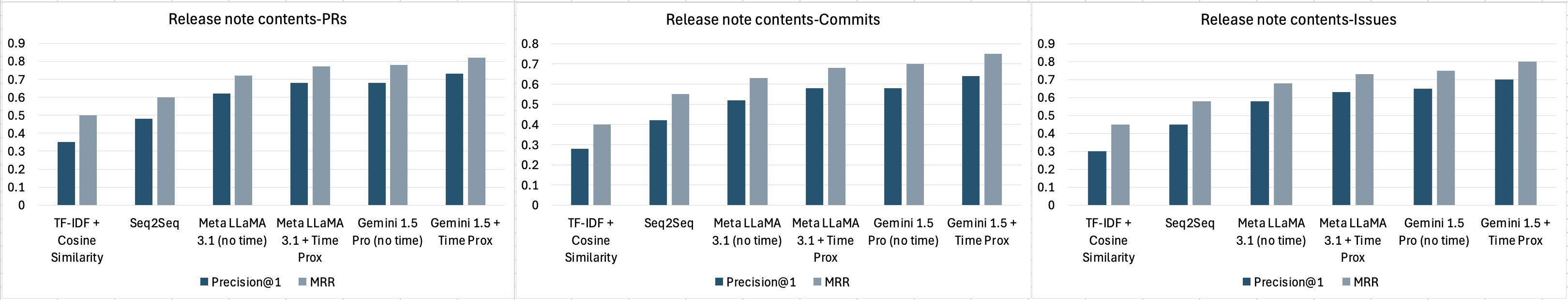}
    \caption{Comparison between our proposed model and the baselines}
    \label{fig:comparison}
\end{figure*}

The evaluation focused on two primary ranking metrics widely used in software traceability research: Precision@1 and MRR. Precision@1 measures the proportion of cases where the top-ranked artifact correctly corresponds to the release content, reflecting the model’s ability to rank the true link at the top. MRR, in contrast, assesses how highly the correct artifact was ranked on average, providing a broader view of ranking effectiveness. These metrics are particularly appropriate for traceability studies because they simulate how developers or tools would prioritize candidate artifacts when attempting to understand or validate traceability links.

Our results showed (Fig. \ref{fig:comparison}) a clear trend: the LLM-based models outperformed traditional baselines across all artifact types. Specifically, the combination of Gemini 1.5 Pro with time proximity achieved the best overall performance. For PR link recovery, Gemini 1.5 Pro with time proximity achieved a Precision@1 of 0.73 and an MRR of 0.82. For commits, the same model achieved 0.64 Precision@1 and 0.75 MRR, while for issues, it reached 0.70 Precision@1 and 0.80 MRR. These figures represent notable improvements over both traditional TF-IDF + cosine similarity methods and Seq2Seq baselines. For example, the TF-IDF approach, which relies solely on text similarity, achieved only 0.35 Precision@1 and 0.50 MRR for PRs. The improvement underscores the advantage of using models that can deeply interpret and reason about textual context.

Traditional methods, such as TF-IDF combined with cosine similarity, performed the weakest across all categories. This result was expected because these approaches rely purely on lexical overlap between release content and artifact text. In many cases, commit messages or PR titles in real-world repositories are too brief or generic to achieve high lexical similarity with release notes. For example, commit messages often contain phrases like ``fix bug" or ``update code", which contribute little discriminative power when compared to more detailed release notes. As a result, TF-IDF struggles to effectively rank the correct links at the top. Additionally, these methods are insensitive to semantic equivalence, meaning they cannot recognize when different words express the same concept.

Seq2Seq models, such as BART-like fine-tuned models, improved upon TF-IDF by learning mappings between release notes and their associated artifacts during training. These models demonstrated better capacity for capturing the structure and phrasing of release content and related PR or commit messages. For instance, the Seq2Seq model achieved 0.48 Precision@1 and 0.60 MRR on PR traceability recovery, outperforming TF-IDF. However, the Seq2Seq model still fell short of LLM-based methods, likely because its training was confined to limited paired data, and it lacked the general reasoning ability and broad knowledge base of LLMs like Meta LLaMA 3.1 or Gemini 1.5 Pro. Furthermore, Seq2Seq models often struggled when faced with variations in wording or less frequent patterns that were underrepresented in the fine-tuning data.

Meta LLaMA 3.1 and Gemini 1.5 Pro demonstrated significant advances in traceability link recovery. These LLMs were queried using few-shot prompting, where they were given examples of correct release-artifact pairs and asked to rank new candidates. Their capacity to model semantic relationships between text segments enabled them to recognize connections between release notes and artifacts even when the exact wording differed. Meta LLaMA 3.1 (without time proximity) achieved Precision@1 values of 0.62, 0.52, and 0.58 for PRs, commits, and issues, respectively, with corresponding MRR values of 0.72, 0.63, and 0.68. Gemini 1.5 Pro (without time proximity) performed slightly better than Meta LLaMA 3.1 across the board, reaching 0.68 Precision@1 for PRs and MRR values of 0.78 for PRs, 0.70 for commits, and 0.75 for issues.

One of the major findings of this study is the contribution of non-textual signals, e.g., time proximity, to improving traceability recovery. When time proximity was integrated into the ranking process, both Meta LLaMA 3.1 and Gemini 1.5 Pro saw noticeable performance boosts. Time proximity was computed by evaluating the temporal distance between the release date and artifact dates (e.g., PR merge date, commit date, issue resolution date), with a closer distance resulting in a higher time score. The final ranking combined the LLM similarity score and the time score using a weighted formula, where the textual reasoning of the LLMs was complemented by the temporal context. This approach is technically important because textual similarity alone can fail in cases where commit messages or PR titles are generic, but time proximity provides an orthogonal signal that helps disambiguate the most likely candidate. The addition of time proximity increased Precision@1 by approximately 5-6\% across the board for both LLMs, and similar gains were observed in MRR. The best results were achieved by Gemini 1.5 Pro with time proximity, highlighting the benefit of combining advanced language understanding with contextual project metadata.

From a technical perspective, Meta LLaMA 3.1 and Gemini 1.5 Pro differ in ways that help explain their relative performance. Meta LLaMA 3.1, accessed through HuggingFace’s API, provided strong reasoning and language understanding capabilities, particularly well-suited to matching patterns between release notes and artifact descriptions. However, its input length and context management capabilities were more constrained than those of Gemini 1.5 Pro. Gemini 1.5 Pro, designed to handle long contexts and multi-part inputs, was better able to process multiple candidates simultaneously, rank them in context, and integrate additional signals like dates. This likely contributed to Gemini 1.5 Pro’s superior performance in this study. Moreover, Gemini's native ranking scores provided richer signals for building combined similarity-time proximity scores, while Meta LLaMA required more explicit post-processing of generation outputs to derive comparable scores.

The findings of this study have important implications for both the research community and practitioners. The superior performance of LLMs, especially when combined with time proximity, demonstrates that traceability link recovery can significantly benefit from approaches that integrate deep semantic reasoning and non-textual project metadata. For open-source projects, this means that automated traceability tools based on LLMs could help contributors, reviewers, and newcomers better navigate the development history, understand how code changes relate to issues and releases, and improve the transparency and maintainability of projects. For researchers, these results highlight the value of combining textual and non-textual features in future traceability work and suggest further opportunities for incorporating other signals, such as contributor identity, file paths, or code structure.

\subsection{RQ3. How do practitioners perceive the impact of the proposed traceability links between release notes and related software artifacts?}

The aim of using a survey study in the traceability link study is to gather practitioners' opinions on the automated traceability links for release notes for improvement in the usability of automated techniques \cite{JSEPSristy}.

\begin{figure}
    \centering
    \includegraphics[width=3.5in]{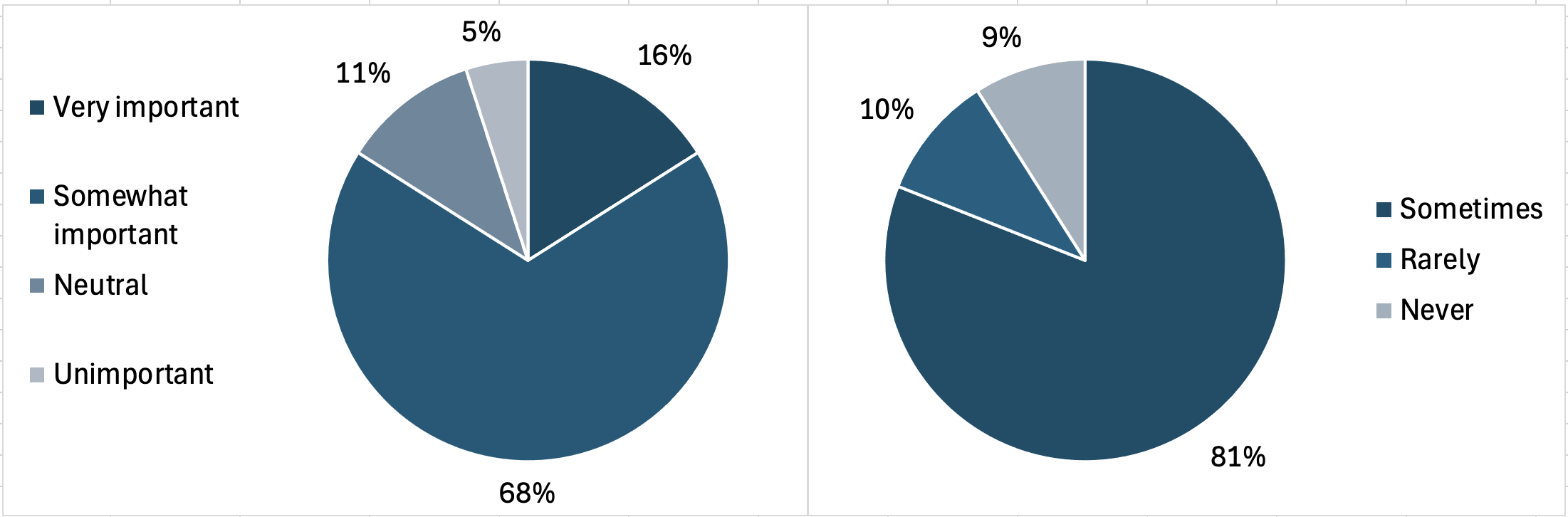}
    \caption{Survey response about a) the importance of traceability maintenance and b) participants maintained traceability in practice}
    \label{fig:surveyresponsecascon}
\end{figure}
During the survey, the participants agreed that PRs, issues, and commits were the most relevant artifacts. A release note summarizes the changes made since the previous release based on the resolved issues and merged pull requests. The contents of a release note may differ depending on the major and minor releases, and setting milestones for each upcoming release can help the release note producers prepare the release notes. 
The participants could not specify the exact time interval or the number of issues resolved during the two releases. However, the release duration could range from 15 days to one month during the product's first year of usage. During this interval, 25-40 issues, pull requests, and 60-70 commits are submitted to the GitHub repository. In the survey, participants were asked about the importance of traceability maintenance, and participants maintained traceability in software development (shown in Fig. \ref{fig:surveyresponsecascon}). Our findings reveal that while 16\% of respondents considered traceability maintenance very important, and 68\% somewhat important, the actual practice of linking artifacts to release notes was inconsistent. A significant proportion of participants (81\%) reported linking artifacts only sometimes, with 10\% rarely and 9\% never doing so.

We summarize the importance of automated traceability links for release notes after getting the response from the participants:

\faComments\ Traceability links greatly aid developers in locating precise source code changes that resolve past release bugs. They streamline the process by bridging commits with release notes, enabling developers to access relevant code changes rapidly. This efficiency saves valuable time and empowers developers to address issues promptly.

\faComments\ Automating the process shows great potential in addressing the missing and broken traceability links. Currently, GitHub contributors laboriously and error-prone insert links to associated artifacts in release notes manually. Adopting an automated traceability approach that can identify and generate these links automatically holds considerable benefits for developers. It reduces the risk of human error, streamlining the creation of accurate and dependable traceability links in release notes.

\faComments\ Traceability links play a pivotal role in boosting collaboration within development teams. These links enable team members to access the context surrounding changes, comprehend their consequences, and collaborate effectively to tackle issues and introduce new features. This level of transparency and accessibility enhances communication and cooperation, resulting in more cohesive and efficient development efforts.

\faComments\ Practitioners identify immense value in traceability links for conducting change impact analysis. These links empower them to assess the consequences of specific code modifications on the software and its functionality. Such insights are pivotal for effective software risk management, allowing practitioners to make informed decisions about code changes and their potential implications on the overall system.

\begin{table}[]
    \centering
    \scriptsize
    \caption{Relevant artifacts based on practitioners' response}
    \label{tab:interviewees_comments}
    \begin{tabular}{p{4.5em}|p{25em}|p{1.5em}}\hline
         \textbf{Artifacts}& \textbf{Comments }&\textbf{\#Res.}  \\ \hline
         \multirow{2}{4.5em}{Milestone} & Milestones help in setting and tracking time-based goals for project completion, such as sprint duration or release dates.& 12 \\\cline{2-3}
         & Software team members can see which issues are part of a milestone for the release and work to achieve those goals. & 7 \\ \hline
         \multirow{2}{4.5em}{README} & According to participants, the README file need to be linked because it provides installation or upgradation instructions of the software to the new release by including system requirements and compatibility information. & 9\\ \cline{2-3}
        & README file often includes usage instructions, showcasing how the software can be used, along with code examples or command-line instructions. & 10 \\ \hline
         Change logs & Changelogs record the technical changes in software development. Developers prefer changelog links with release notes to understand the software's evolution from previous releases and track issues or enhancements over time. Changelogs focus on listing changes made to the source code. & 18\\ \hline
         \multirow{2}{4.5em}{Migration guide} & A migration guide helps ensure a smooth transition for software practitioners. It provides information on the compatibility of the new release with the previous version and any potential issues that users might encounter when migrating. & 10 \\\cline{2-3}
         & For applications with data storage, the guide explains how to migrate data from the old version to the new one, ensuring that no data is lost or corrupted during the process. & 5\\ \hline
         \multirow{2}{4.5em}{API documenta-tion} &  API documentation highlights the modifications and additions of API endpoints, parameters, or data structures& 11 \\\cline{2-3}
         & API changes can have security implications. Practitioners need to be aware of these changes to maintain the security of their systems and data. & 7\\\hline
         Support Resources & Participants agree to include links to support channels with release notes, such as FAQs, user community forums, contributors list or contact information for technical assistance. & 13 \\\hline
    \end{tabular}
\end{table}

Additionally, we asked the practitioners about enhancing the release notes traceability links study. We try to understand the real-life release notes practice from practitioners' responses. Table \ref{tab:interviewees_comments} represents the other relevant artifacts of release notes that need to be linked according to practitioners' aspects. We summarize the reasons why participants mention these relevant artifacts are important in the survey study and provide the number of responses.

Participants were informed that release notes not only contain the software change information for users but also include the links to the installation process and configuration information. Readme files and the upgradation guide are necessary to link with the release notes. In this study, one project lead said that at a high level of abstraction, release notes summarize the key features, enhancements, and bug fixes in a particular software release with user-friendly language; however, keeping track of technical changes, e.g., source code modification, is important for technical users. Therefore, change logs and API documentation are other relevant artifacts of release notes in practice.
The requirements incorporated in releases determine what users can get from the new releases. Release notes contain rich information about the new and implemented requirements of the projects. Participants mentioned that there exists a close relationship between information documented in release notes and defined milestones of projects. For example, a set of software requirements may be enclosed in the upcoming release, but it is hardly tracked with the release. Two QA engineers and five developers described the importance of keeping track of the milestone with the release, as testers can know the overall milestone and what is tested, or which requirements need to be tested.

Moreover, thirteen participants mentioned supporting resources of release notes. Survey participants described that linking to FAQs and user community forums provides users with a valuable resource for finding answers to common questions, troubleshooting issues, and accessing peer-to-peer support. These artifacts enable practitioners to find solutions quickly, reducing the burden on the support team and leading to faster issue resolution and lower support costs. Including contributors' names in release notes is important for open-source projects because it ensures accountability of the development process and helps in community building, and form of promotion and showcases their skills and contributions.

\section{Implications}\label{sec:discussion}
\paragraph{For Contributors}
The proposed traceability approach helps contributors better understand how their changes, across issues, PRs, and commits, connect to release notes. This transparency supports contributors in documenting and communicating their work more effectively, ensuring that their contributions are accurately reflected in project history. It also enables contributors to demonstrate their impact on the project, which can be valuable for recognition within the community or during performance evaluations in industry settings.

\paragraph{For Reviewers}
The approach provides reviewers and maintainers with a systematic way to trace and supervise code contributions linked to specific issues and features. By recovering traceability links automatically, reviewers can more easily verify whether commits and PRs properly address the related issues and whether they have been correctly reflected in the release notes. This can reduce oversight during code review and help identify missing documentation or incomplete fixes that might otherwise introduce technical debt.

\paragraph{For Newcomers}
New contributors often face challenges understanding how code changes flow through PRs, commits, and issues, and how these changes are ultimately incorporated into releases. Our approach assists newcomers in learning the code review process by observing how previous changes were reviewed, merged, and documented.
It helps in identifying similar prior changes linked to similar issues, providing valuable examples to follow and understanding the project's maintenance practices through clear, recovered links between code artifacts and releases, helping them contribute more effectively and confidently.

By improving the recovery of traceability links, this work helps enhance the maintainability and usability of open-source projects. Clear traceability supports better change impact analysis, reduces the risk of undocumented changes, and makes it easier for contributors and maintainers to track the evolution of the project over time. It can also help identify opportunities for code reuse or guide refactoring efforts by showing how similar issues were addressed in the past.

\section{Threats to Validity}\label{sec:threats}
External Validity refers to how broadly our analysis can be generalized. Since our analysis focuses on open-source repositories, it may not fully reflect the traceability needs and practices of enterprise-level projects, which often have different workflows, release processes, and documentation practices. While we believe that industry projects typically have structured expert-level communication and knowledge-sharing processes, open-source projects, especially for new contributors, often lack formal communication channels. Therefore, our study specifically focuses on these open-source contributors.
A potential external validity threat is the use of external issue-tracking systems, such as JIRA, to manage issues and track changes. In some projects, issues, PRs, and commits may be linked through external systems rather than solely within GitHub. However, JIRA also maintains issue titles, statuses, and merge information. If such data is included in the dataset, the proposed model, which relies on text similarity and time proximity, could still effectively find the pairs. Future work could explore adapting the model to handle multi-platform traceability data.

Threats to Internal Validity pertain to dataset preparation. We considered a single pair of release note contents linked to an issue, commit, and PR. However, in practice, multiple commits may be required to resolve an issue. Although the proposed model considers both text similarity and time proximity, allowing it to track multiple commits, this aspect could be further expanded in future work.

\section{Related Work}\label{sec:related}
Several studies have proposed automated approaches for recovering traceability links between {source code - documentation} \cite{icseTLRsourceandcode, linkcodeDoc}, {issues - pull requests} \cite{pilinkissueprlinkML}, {bug - source code changes} \cite{relinkbugsandchangesFSE} and {commits - issues} \cite{frlink, BTLinkissuecommit}. 

Antoniol et al. \cite{linkcodeanddoc} proposed an information retrieval-based method to recover traceability links between source code and free text documents as well as discussed some notable examples of the usefulness of this technique in software maintenance and requirement tracing. The paper discussed the differences between two IR models, and found the probabilistic model achieves the highest recall values with a smaller number of documents, and the vector space model shows regular progress in the recall values when increasing the number of documents retrieved. But, hybrid \cite{codeanddocase2011} and deep-learning based \cite{isuecommitdeeplinkjss2019} techniques predict more accurate traceability links.

Le et al. \cite{rclinkerissuecommiticpc} developed a solution for recovering traceability links between bug reports and commits by leveraging cosine similarity between the entire issue report (i.e., summary, description, and all comments) and the commit message, as well as the average and maximum cosine similarity between different textual elements of the issue report and the commit. 
The limitation is if the bug reports and commits have low similarity, this approach may not be able to recover the missing links.
Wu et al. \cite{relinkbugsandchangesFSE} proposed a ReLink approach to recover missing links between bugs and committed changes and  presented significantly better accuracy (89\% precision and 78\% recall on average) than those of traditional heuristic approaches. This study focused on the time duration  when a bug is created and when the corresponding commits are submitted, without considering whether these commits have been merged into the main branch or not. 

DeepLink \cite{deeplinkissuecommitsaner2019} resolved the issue–commit link recovery problem and considered the code knowledge graph to incorporate the semantics of code context into the issue–commit link recovery model and thus improve its performance. This study embedded the source code to calculate semantic similarity and code similarity using a deep learning approach before training an SVM binary classification model. 
PULink \cite{issuecommitASE2017} approach to recover the links between issue reports and corresponding fix commits by using metadata features, e.g., issue type, commit type, interval, and interval type, and similarity features, e.g., code feature similarity and text feature similarity. 
Alshara et al. \cite{pilinkissueprlinkDS} aimed to examine the association rules between issues and their corresponding linked PRs and considered textual similarities as the features. Our study takes into account both textual and non-textual data such as similarity scores, version numbers, closed dates, and status components to predict traceability links. Including non-textual data can enhance the prediction of related artifacts in cases where text similarity is low.

Several additional studies have been conducted to establish benchmarks for the recovery of traceability links. Fuchss et al. \cite{benchmarkarchimodel} developed an open-source benchmark dataset to set clear standards and baseline results for traceability link recovery between software architecture documentation and software architecture models.
Similarly, Alshara et al. \cite{pilinkissueprlinkDS} proposed a metamodel based on Issues and Pull Requests from Android projects on GitHub. They examined the relationships between Issues and their linked Pull Requests by extracting four features from the titles, bodies, labels, and comments. Additionally, they calculated similarities based on three lexical and one semantic similarity metric, considering both the lengths and content of the features.

\section{Conclusion}\label{sec:conclusion}
This study demonstrated the effectiveness of integrating both textual and non-textual features, specifically time proximity, in improving the recovery of traceability links between release notes and the corresponding PRs, commits, and issues using a fine-tuned LLM-based approach. For future work, tracking the backporting of issues \cite{Backports}, PRs, and commits across different software versions holds a promising research idea that maintains long-term support for previous releases or needs to address security patches in older versions.

\section*{Acknowledgment}
This research is supported in part by the Natural Sciences and Engineering Research Council of Canada (NSERC) Discovery Grants program, the Canada Foundation for Innovation's John R. Evans Leaders Fund (CFI-JELF), and by the industry-stream NSERC CREATE in Software Analytics Research (SOAR).
\bibliographystyle{plain}
\bibliography{ref}
\end{document}